\documentclass[a4paper,10pt,twoside]{cpc-hepnp}
\usepackage{CJK,upgreek,fancyhdr}
\usepackage{multicol}
\usepackage{graphicx}
\usepackage{booktabs}
\usepackage{amssymb,bm,mathrsfs,bbm,amscd}
\usepackage[tbtags]{amsmath}
\usepackage{lastpage}
\usepackage[english]{babel}
\usepackage{xcolor}

\begin{document}

\fancyhead[c]{\small Chinese Physics C~~~Vol. xx, No. x (201x) xxxxxx}
\fancyfoot[C]{\small 010201-\thepage}

\footnotetext[0]{Received 31 June 2015}

\title{Local probes strongly favor $\Lambda$CDM against power-law and $R_h=ct$ universe\thanks{This work has been supported by the National Natural Science Fund of China under grant Nos. 11603005, 11775038, 11647307, 11675182 and 11690022.}}

\author{%
      Hai-Nan Lin$^{1;1)}$\email{linhn@cqu.edu.cn}%
\quad Xin Li$^{1;2)}$\email{lixin1981@cqu.edu.cn}%
\quad Yu Sang$^{2,3;3)}$\email{sangyu@ihep.ac.cn}
}
\maketitle

\address{%
$^{1}$Department of Physics, Chongqing University, Chongqing 401331, China\\
$^{2}$Institute of High Energy Physics, Chinese Academy of Sciences, Beijing 100049, China\\
$^{3}$University of Chinese Academy of Sciences, Beijing 100049, China\\
}

\begin{abstract}
  We constrain three cosmological models, i.e. the concordance cold dark matter plus a cosmological constant ($\Lambda$CDM) model, Power-law (PL) model, and $R_h=ct$ model using the available local probes, which includes the JLA compilation of type-Ia supernovae (SNe Ia), the direct measurement of Hubble constant (H(z)), and the baryon acoustic oscillations (BAO). For $\Lambda$CDM model, we consider two different cases, i.e. zero and non-zero spatial curvature. We find that by using the JLA alone, it is indistinguishable between $\Lambda$CDM and PL models, but the $R_h=ct$ model is strongly disfavored. If we combine JLA+H(z), the $\Lambda$CDM model is strongly favored against the other two models. The combination of all the three datasets also supports $\Lambda$CDM as the best model. We also use the low-redshift ($z<0.2$) data to constrain the deceleration parameter using cosmography method, and find that only the $\Lambda$CDM model is consistent with cosmography. However, there is no strong evidence to distinguish between flat and non-flat $\Lambda$CDM models by using the local data alone.
\end{abstract}



\footnotetext[0]{\hspace*{-3mm}\raisebox{0.3ex}{$\scriptstyle\copyright$}2013
Chinese Physical Society and the Institute of High Energy Physics
of the Chinese Academy of Sciences and the Institute
of Modern Physics of the Chinese Academy of Sciences and IOP Publishing Ltd}%

\begin{multicols}{2}

\section{Introduction}\label{sec:introduction}

The progresses on both the experiments and theories of cosmology in recent decades lead to the foundation of standard model, i.e. the  cold dark matter plus a cosmological constant ($\Lambda$CDM) model. According to this model, the universe mainly consists of cold dark matter and dark energy (cosmological constant), while the ordinary baryonic matter only occupies a small proportion of the total contents. The cosmological constant is responsible for the accelerated expansion of the universe, which was first discovered from the fact that the luminosity of type-Ia supernovae (SNe Ia) is dimmer than expected \cite{Perlmutter:1999,Riess:1998}. The $\Lambda$CDM model is well consistent with various local observations such as SNe Ia, the direct measurement of Hubble parameters (H(z)), and the baryon acoustic oscillations (BAO). More importantly, it is consistent with the cosmic microwave background from the WAMP \cite{Bennett:2011,Bennett:2013} and Planck satellites \cite{Ade:2013zuv,Ade:2015xua}.

Although the $\Lambda$CDM model achieves great successes, it also confronts many challenges, among which the most famous ones are the ``cosmological constant fine tuning problem" and the ``cosmic coincidence problem" \cite{Weinberg:1989,Zlatev:1998tr}. The former asks why the cosmological constant is so close to zero but not exactly zero, and the latter concerns why the densities of dark matter and dark energy approximately equal to each other today. In addition, it is found that the Hubble constant measured from the local SNe Ia and Cepheids is in more than $3\sigma$ tension from that obtained from the CMB \cite{Riess:2016jrr}. These problems motivate cosmologists to pursue new theories beyond the standard model.

Another problem $\Lambda$CDM confronts is the horizon problem, which asks why the universe appears statistically homogeneous and isotropic in accordance with the cosmological principle. According to the standard big bang model, the gravitational expansion dost not allows the universe to reach to the thermal equilibrium, hence it is difficult to explain the homogeneity and anisotropy. Although the horizon problem can be solved by adding an exponential inflation epoch to the very early of the universe, another problem inevitable arises. That is, why the gravitational horizon equals to the distance light has traveled since the big bang at current epoch. According to $\Lambda$CDM, there is only one time at which the gravitational horizon equals to the light travelling distance \cite{Melia:2003}. It is difficult to explain why this equality happens exactly at present day, but not at early or later time. To avoid this coincidence, Melia \cite{Melia:2007} proposed the $R_h=ct$ model, in which the gravitational horizon always equal to the light travelling distance throughout the whole history of universe. It was showed that various local data are well consistent with the $R_h=ct$ model \cite{Melia:2013hf,Wei:2015dsa,Melia:2015nwa,Wei:2016jqa,Melia:2017djn}. A detailed analysis on the combined data of local probes, however, showed that the $R_h=ct$ model is strongly disfavored \cite{Bilicki:2012ub,Shafer:2015kda,Haridasu:2017lma}.

An alternative model more general than $R_h=ct$ is the power-law (PL) model \cite{Dolgov:1997,Dolgov:2014faa}, which assumes that the universe expands in a simple power law, i.e. the scale factor of the universe follows $a(t)\propto t^n$. Although it is unlikely that PL model can describe the whole evolution history of the universe, some investigations showed that it is consistent with various low-redshift data \cite{Dolgov:2014faa,Sethi:2005au,Zhu:2007tm}. Especially, it was showed that PL model with index $n\sim 1.5$ can fit the SNe Ia data as well as $\Lambda$CDM model \cite{Tutusaus:2016nml}. On the other hand, the validity of PL model is also questioned by some authors \cite{Shafer:2015kda,Haridasu:2017lma}.

One of the most important discoveries in modern cosmology is the accelerated expansion of the Universe. This phenomenon was first discovered from observation on the luminosity of SNe Ia in the end of 1990s, which was latter awarded the Nobel Prizes \cite{Perlmutter:1999,Riess:1998}. Nowadays the acceleration of the Universe and the existence of dark energy are widely accepted by cosmologists. Recently, however, some investigations showed that the evidence for acceleration can be weaken. By using unconventional priors on the SN parameters, Nielsen et al. \cite{Nielsen:2016pga} found that the SNe Ia data are still quite consistent with a constant rate of expansion. Tutusaus et al. \cite{Tutusaus:2017ibk} found that the non-accelerated power-law model is a good fit to various local data if the cosmological evolution of the intrinsic luminosity of SNe is taken into account. A model-independent way to test the acceleration of the Universe is using cosmography method. We note that the $R_h=ct$ model is a non-accelerating model, thus if the Universe is proven to be accelerating, then $R_h=ct$ model can be ruled out.

In this paper, we use various local probes, which including the SNe Ia, H(z) and BAO, to test three cosmological models, i.e. $\Lambda$CDM model, PL model and $R_h=ct$ model. To avoid the model-dependence, the cosmography method is also used to constrain the deceleration parameter. The rest of the paper is organized as follows: In section 2, we briefly review the cosmological models. In section 3, we introduce the observational datasets that are used to constrain the cosmological models. In section 4, we use the Markov chain Monte Carlo method to calculate the posterior probability density function of cosmological parameters, and then use the information criteria to pick up the model which is best consistent with the data. Finally, discussions and conclusions are given in section 5.

\section{Cosmological models}\label{sec:models}

In this section, we briefly review three cosmological models we are interested in, including the $\Lambda$CDM, PL and $R_h=ct$ models.

The $\Lambda$CDM model is the standard models and was proven to be consistent with various observations. It is based on the homogeneous and isotropic Friedmann-Robertson-Walker (FRW) metric
\begin{equation}
  ds^2=c^2dt^2-a^2(t)\left(\frac{dr}{1-kr^2}+r^2d\theta^2+r^2\sin^2\theta d\phi^2\right),
\end{equation}
where $a(t)$ is the scale factor, and $k=0,\pm 1$ is the curvature parameter of the universe. Substituting the FRW metric into the Einstein field equations results to the Friedmann equation
\begin{equation}\label{eq:friedmann}
  \left(\frac{\dot{a}}{a}\right)^2=\frac{8\pi G\rho}{3c^2}-\frac{kc^2}{a^2}
\end{equation}
and the acceleration equation
\begin{equation}\label{eq:acceleration}
  \frac{\ddot{a}}{a}=-\frac{4\pi G}{3c^2}(\rho+3p),
\end{equation}
where $\rho=\rho_r+\rho_m+\rho_\Lambda$ is the total energy density of the universe, which includes the radiation, matter and dark energy. Assuming the equations of state (EoS) $w\equiv p/\rho$ for the radiation and matter components equal to $1/3$ and $0$ respectively, we obtain that $\rho_r$ scales as $a^{-4}$ and $\rho_m$ scales as $a^{-3}$. We further assume that the dark energy is a constant and does not evolve with $a$.

Defining the Hubble parameter $H=\dot{a}/a$ and the critical energy density $\rho_{c,0}=3c^2H_0^2/8\pi G$, the Friedmann equation (\ref{eq:friedmann}) can be rewritten as
\begin{equation}
  \left(\frac{H}{H_0}\right)^2=\Omega_r a^{-4}+\Omega_m a^{-3}+\Omega_k a^{-2}+\Omega_\Lambda,
\end{equation}
where $\Omega_i\equiv \rho_{0,i}/\rho_{c,0}$ $(i=r,m,\Lambda)$ is the normalized energy density today, $\Omega_k\equiv -kc^2/H_0^2$, and $H_0$ is called the Hubble constant. The total energy density is normalized to unity, i.e. $\Omega_r+\Omega_m+\Omega_k+\Omega_\Lambda=1$. Using the relation $a=1/(1+z)$, the Hubble parameter can be rewritten as a function of redshift,
\begin{equation}
  H(z)=H_0\sqrt{\Omega_r(1+z)^4+\Omega_m(1+z)^3+\Omega_k(1+z)^2+\Omega_{\Lambda}}.
\end{equation}
Another important quantity is the deceleration parameter, which is defined by $q=-\ddot{a}a/\dot{a}^2$. A positive or negative $q$ means that the universe is decelerating or accelerating. From equations (\ref{eq:friedmann}) and (\ref{eq:acceleration}) the deceleration parameter can be written as a function of the mass components of the universe
\begin{equation}
  q_0=\Omega_r+\frac{1}{2}\Omega_m-\Omega_{\Lambda}.
\end{equation}
At present day the radiation component is negligible compared to the rest components, so we fixed $\Omega_r=0$. If the universe is spatially flat, i.e. $\Omega_k= 0$, the deceleration parameter only depends on the energy density of matter, $q=(3/2)\Omega_m-1$. Such a model is the so-called concordance cosmological model. Here we consider the flat and non-flat $\Lambda$CDM models separately.
The comoving distance is given by \cite{Hogg:1999ad}
\begin{equation}
  D_C=\frac{c}{H_0}\int_0^z\frac{dz}{\sqrt{\Omega_m(1+z)^3+\Omega_k(1+z)^2+\Omega_\Lambda}}
\end{equation}
The luminosity distance is related to the comoving distance by
\begin{eqnarray}
  D_L=\begin{cases}(1+z)\frac{c}{H_0}\frac{1}{\sqrt{\Omega_k}}\sinh\left(\sqrt{\Omega_k}D_CH_0/c\right), & \Omega_k>0,\\
  (1+z)D_C, & \Omega_k=0,\\
  (1+z)\frac{c}{H_0}\frac{1}{\sqrt{-\Omega_k}}\sin\left(\sqrt{-\Omega_k}D_CH_0/c\right), & \Omega_k<0.\\
  \end{cases}
\end{eqnarray}

The power-law model \cite{Dolgov:1997,Dolgov:2014faa} is a toy model and is based on the assumption that the scale factor of the universe expands as a simple power law, namely $a(t)=(t/t_0)^n$, in regardless of the contents of the universe, where $t_0$ is the current age of the Universe. In the power-law model, the Hubble parameter reads
\begin{equation}
  H(z)=H_0(1+z)^{\frac{1}{n}}.
\end{equation}
The deceleration parameter is given by $q=1/n-1$, and $n>1$ or $n<1$ means an accelerating or a decelerating universe, respectively.

The $R_h=ct$ universe \cite{Melia:2007,Melia:2011fj} is based on the assumption that the gravitational horizon $R_h$ equal to the distance $ct$ light has traveled since the big bang all the way the cosmos expansion. In $R_h=ct$ model, the universe also consists of radiation, matter and dark energy, as the $\Lambda$CDM does. The main difference between $R_h=ct$ and $\Lambda$CDM is that the former has no assumption on the EoS of dark energy but requires that the EoS of the total contents to be $w\equiv p/\rho=-1/3$. According to the $R_h=ct$ universe, the Hubble parameter is given by
\begin{equation}
  H(z)=H_0(1+z).
\end{equation}
In this model, the universe expands steadily and the deceleration parameter is zero.

In the PL and $R_h=ct$ models, the luminosity distance is given by $D_L=(1+z)c\int_0^z[1/H(z)]dz$. Therefore, we have
\begin{eqnarray}
  D_L=\begin{cases}
  (1+z)\frac{c}{H_0}\frac{(1+z)^{1-\frac{1}{n}}-1}{1-\frac{1}{n}}, & {\rm PL},n\neq 1,\\
  (1+z)\frac{c}{H_0}\ln(1+z), & {\rm PL},n=1~\&~R_h=ct.
  \end{cases}
\end{eqnarray}
It is convenient to convert the luminosity distance to the dimensionless distance modulus by
\begin{equation}
  \mu=5\log\frac{D_L}{{\rm Mpc}}+25,
\end{equation}
where ``\,log" represents the logarithm of base 10.

One of the model-independent ways to describe the local Universe is the so-called cosmography \cite{Dunsby:2016ers}. The main idea of cosmography is to expand the scale factor $a(t)$ and other quantities of interests into Taylor series. In this way the Hubble parameter reads
\begin{equation}
  H(z)=H_0[1+(1+q_0)z+\mathcal{O}(z^2)],
\end{equation}
where $H_0$ is the Hubble constant and $q_0$ is the deceleration parameter at present day. The luminosity distance is given by
\begin{equation}
  D_L(z)=\frac{cz}{H_0}\left[1+\frac{1}{2}(1-q_0)z+\mathcal{O}(z^2)\right].
\end{equation}
The cosmography only valid when $z\ll 1$.

\section{Data and methodology}\label{sec:data}

In this section, we use the available local data to constrain the cosmological models. These local data include SNe Ia, H(z) and BAO.

The first local probes used in our paper are SNe Ia. Due to the approximately constant absolute luminosity, SNe Ia are widely used as the standard candles to constrain the cosmological parameters. Recently, many SNe Ia samples are released \cite{Kowalski:2008ez,Amanullah:2010,Suzuki:2012,Betoule:2014frx}. Here we use the most up-to-date compilation of SNe Ia, i.e. the JLA sample \cite{Betoule:2014frx}. The JLA consists of 740 SNe Ia in the redshift range $[0.01,1.30]$. Each SN has well measured light curve parameters. The distance moduli of SNe can be extracted from the light curves using the empirical relation \cite{Betoule:2014frx,Tripp:1998,Guy:2005}
\begin{equation}\label{eq:mu_sn}
  \hat\mu=m_B^*-(M_B+\delta\cdot\Delta_M-\alpha X_1+\beta \mathcal{C}).
\end{equation}
where $m_B^*$ is the observed peak magnitude, $M_B$ is the absolute magnitude, $X_1$ is the stretch factor, and $\mathcal{C}$ is the supernova color at maximum brightness. Strictly speaking, the absolute magnitude is not a constant, and it depends on the host galaxy complexly. Following Ref.\cite{Betoule:2014frx}, we use a simple step function to approximate such a dependence, i.e., we add a term $\delta\cdot\Delta_M$ to $M_B$ and set $\delta=1$ (or $\delta=0$) if the mass of host galaxy is larger (or smaller) than $10^{10}M_{\odot}$. The two parameters $\alpha$ and $\beta$ are universal constants and they can be fitted simultaneously with cosmological parameters. The best-fitting parameters are the ones which can maximize the likelihood
\begin{equation}\label{eq:likelihood_sn}
\mathcal{L}_{\rm SN}= \frac{1}{\sqrt{\det{(2\pi \bm{C})}}}\exp\left[-\frac{1}{2}(\bm{\mu}-\bm{\hat{\mu}})\dag \bm{C}^{-1}(\bm{\mu}-\bm{\hat{\mu}})\right],
\end{equation}
where $\bm C$ is the covariance matrix of $\bm{\hat{\mu}}$. Note that $\bm C$ not only depends on the light curve parameters, but also depends on the nuisance parameters $\alpha$ and $\beta$. Therefore, the normalization factor in equation (\ref{eq:likelihood_sn}) is not a constant and couldn't be neglected. In each iteration of the minimization procedure, $\bm C$ should be recalculated using the new parameters. The detailed information on the covariance matrix can be found in Ref.\cite{Betoule:2014frx}.

The second local probes used here are H(z) data, which directly measure the Hubble parameter at different redshift. Two commonly used methods to measure H(z) are the differential age of galaxies (DAG) method \cite{Jimenez:2002gg,Stern:2010,Moresco:2012} and the BAO method \cite{Gaztanaga:2008xz,Blake:2012}. The DAG method measures H(z) by comparing the age of galaxy at different redshift, and the BAO method extract H(z) from the peak of acoustic oscillation of baryon. The H(z) data have the advantage over SNe because the latter rely on the integral of the cosmic expansion history rather than the expansion history itself. After the integration some important information may be erased out. However, the H(z) data from BAO method is more or less model dependent, only the DAG method is free of cosmological model. In this paper, we use the 30 H(z) data obtained using the DAG method compiled in Ref.\cite{Moresco:2016mzx}. The likelihood for H(z) data is given by
\begin{equation}
  \mathcal{L}_{\rm H}\propto \exp\left[-\frac{1}{2}\sum_{i=1}^N\frac{[H(z_i)-\hat{H}_i]^2}{\sigma_{\hat{H}_i}^2}\right],
\end{equation}
where $H(z_i)$ is the theoretical Hubble parameter at redshift $z_i$, $\hat{H}_i$ is the observed Hubble parameter, and $\sigma_{\hat{H}_i}$ is the uncertainty of $\hat{H}_i$.

The final local probes used in our paper are the BAO data. BAO are regular, periodic fluctuations in the density of the visible baryonic matter of the universe. As the SNe Ia provide a ``standard candle" for astronomical observations, BAO provides a ``standard ruler" for length scale in cosmology. This standard ruler is characterized by the sound horizon $r_d$ when the baryons decoupled from the Compton drag of photons at redshift $z_d$ \cite{Eisenstein:1998},
\begin{equation}\label{eq:sound-horizon}
  r_d=\int_{z_d}^{\infty}\frac{c_s(z)dz}{H(z)},
\end{equation}
where $c_s(z)$ is the sound speed at redshift $z$. The value of $r_d$ strongly depends on the early epoch of the universe, and different models may have very different $r_d$. From the local data alone we couldn't get information about the early universe. Following Ref. \cite{Tutusaus:2017ibk}, we treat $r_d$ as a free parameter.

BAO measure the ratio of the effective distance to the sound horizon, i.e. $R(z)=D_V(z)/r_d$, where
\begin{equation}
  D_V(z)=\left[\frac{d_L^2(z)}{(1+z)^2}\frac{cz}{H(z)}\right]
\end{equation}
is the effective distance, which takes into consideration the anisotropic expansion in radial and transverse direction. In this paper, we use the seven BAO data points compiled in Table 1 of Ref.\cite{Cheng:2014kja}. These data are the compilation of BAO from the 6dF Galaxy Survey \cite{Beutler:2011}, Baryon Oscillation Spectroscopic Survey \cite{Anderson:2014,Delubac:2014aqe}, and the WiggleZ Dark Energy Survey \cite{Kazin:2014qga}, The likelihood of BAO data is given by
\begin{equation}
  \mathcal{L}_{\rm BAO}\propto\exp\left[-\frac{1}{2}\sum_{i=1}^N\frac{[R_{\rm th}(z_i)-R_{\rm obs}(z_i)]^2}{\sigma_{R_i}^2}\right].
\end{equation}

Finally, we combine all the data sets to constrain the cosmological models. The total likelihood of the combined data sets is the product of the individual likelihoods, i.e.
\begin{equation}
  \mathcal{L}_{\rm total}=\mathcal{L}_{\rm SN}\cdot\mathcal{L}_{\rm H}\cdot\mathcal{L}_{\rm BAO}.
\end{equation}

We use the information criteria (IC) to pick up the model which can best depict the data. The two most widely used IC are the Akaike information criterion (AIC) \cite{Akaike:1974} and the Bayesian information criterion (BIC) \cite{Schwarz:1978}. They are defined by
\begin{equation}
  {\rm AIC}=-2\ln\mathcal{L}_{\rm max}+2k,
\end{equation}
\begin{equation}
  {\rm BIC}=-2\ln\mathcal{L}_{\rm max}+k\ln N,
\end{equation}
where $\mathcal{L}_{\rm max}$ is the maximum of likelihood, $k$ is the number of free parameters, and $N$ is the number of data points. The model which has the smallest IC is the best one. It is not the absolute value of IC but the difference of IC between different models that are important in the model comparison. We use the flat $\Lambda$CDM as the fiducial model, and define the difference of IC of a model with respect to that of flat $\Lambda$CDM as
\begin{equation}
  \Delta {\rm IC}_{\rm model}={\rm IC}_{\rm model}-{\rm IC}_{\rm flat-\Lambda CDM}.
\end{equation}
According to the Jeffreys' scale \cite{Jeffrey:1998,Liddle:2007}, a model with $\Delta {\rm IC}>5$ or $\Delta {\rm IC}>10$ means that there is `strongly' or `decisive' evidence against this model with respect to the flat $\Lambda$CDM.

\section{Results}\label{sec:results}

We use the publicly available python package \textsf{emcee} \cite{ForemanMackey:2012ig} to calculate the posterior probability distribution functions of free parameters. A flat prior is used on each parameter. First, we use the JLA data alone to constrain the cosmological parameters. In this case, The Hubble constant $h_0$ ($h_0=H_0/100~{\rm km}~{\rm s}^{-1}~{\rm Mpc}^{-1}$) is degenerated with the absolute magnitude $M_B$, so they couldn't be constrained simultaneously. Therefore, we fix $h_0=0.7$ and leave $M_B$ free. The mean and $1\sigma$ error of each parameter are reported in Table \ref{tab:parameter1}. In the last three rows, we also report the $\ln\mathcal{L}_{\rm max}$, $\Delta{\rm AIC}$ and $\Delta{\rm BIC}$. According to the IC, there is decisive evidence against the $R_h=ct$ model. However, it is indistinguishable between flat $\Lambda$CDM and PL models using SNe data alone. According to AIC, the flat and non-flat $\Lambda$CDM models fit the data equally well, while according to BIC, the data favors flat $\Lambda$CDM model. In the non-flat $\Lambda$CDM model, the $\Omega_m$ and $\Omega_\Lambda$ values are somewhat smaller than, but are still marginally consistent with the Planck 2015 results \citep{Ade:2015xua} within $1\sigma$ uncertainty. In the PL model, $n>1$ means that JLA data favors an accelerating universe.
\end{multicols}
\begin{table}[htbp]
\centering
\caption{\small{The best-fitting parameters and their $1\sigma$ uncertainties from JLA.}}
\begin{tabular}{crrrr}
  \hline
  & flat $\Lambda$CDM & non-flat $\Lambda$CDM  & PL & $R_h=ct$ \\
  \hline
  $\Omega_m$ & $0.329\pm 0.034$ & $0.226\pm 0.105$ & \---- & \---- \\
  $\Omega_\Lambda$ & \---- & $0.522\pm 0.163$ & \---- & \---- \\
  $n$ & \---- & \---- & $1.421\pm 0.116$ & \---- \\
  $\alpha$ & $0.127\pm 0.006$ & $0.127\pm 0.006$ & $0.126\pm 0.006$ & $0.124\pm 0.006$ \\
  $\beta$ & $2.633\pm 0.066$ & $2.627\pm 0.067$ & $2.631\pm 0.066$ & $2.606\pm 0.067$ \\
  $M_B$ & $-19.053\pm 0.023$ & $-19.044\pm 0.025$ & $-19.030\pm 0.024$ & $-18.946\pm 0.017$ \\
  $\Delta_M$ & $-0.054\pm 0.022$ & $-0.054\pm 0.021$ & $-0.056\pm 0.023$ & $-0.059\pm 0.022$ \\
  $\ln \mathcal{L}_{\rm max}$ & $337.873$ & $338.509$ & $338.188$ & $322.076$\\
  $\Delta {\rm AIC}$ & 0 & 0.728 & -0.630 & 29.594 \\
  $\Delta {\rm BIC}$ & 0 & 5.335 & -0.630 & 24.987\\
  \hline
\end{tabular}\label{tab:parameter1}
\end{table}
\begin{multicols}{2}

Then we add H(z) data to the JLA and make a combined analysis. Adding H(z) data breaks the degeneracy between $H_0$ and $M_B$, so they can be fitted simultaneously. The besting-fitting parameters are listed in Table \ref{tab:parameter2}. In the last two rows we also list the $\Delta{\rm AIC}$ and $\Delta{\rm BIC}$ values for each model. We may see that there is decisive evidence against PL and $R_h=ct$ models. However, there is weak or strong evidence favoring flat-$\Lambda$CDM against non-flat $\Lambda$CDM, based on whether AIC or BIC criterion is chosen. In the PL model, the besting-fitting power-law index $n$ is reduced compared wih the JLA only case. In the $\Lambda$CDM (both flat and non-flat) models and PL models, the Hubble constant is more consistent with that of Planck 2015 results \cite{Ade:2015xua} than to the local value from Cepheids \citep{Riess:2016jrr}. However, in the $R_h=ct$ model, the Hubble constant is unexpectedly small.
\end{multicols}
\begin{table}[htbp]
\centering
\caption{\small{The best-fitting parameters and their $1\sigma$ uncertainties from JLA+H(z).}}
\begin{tabular}{crrrr}
  \hline
  & flat $\Lambda$CDM & non-flat $\Lambda$CDM & PL & $R_h=ct$ \\
  \hline
  $h_0$ & $0.677\pm 0.019$ & $0.671\pm 0.023$ & $0.683\pm 0.019$ & $0.625\pm 0.014$ \\
  $\Omega_m$ & $0.324\pm 0.028$ & $0.245\pm 0.095$ & \---- & \---- \\
  $\Omega_\Lambda$ & \---- & $0.542\pm 0.157$ & \---- & \---- \\
  $n$ & \---- & \---- & $1.289\pm 0.071$ & \---- \\
  $\alpha$ & $0.128\pm 0.006$ & $0.127\pm 0.006$ & $0.126\pm 0.006$ & $0.124\pm 0.006$ \\
  $\beta$ & $2.631\pm 0.066$ & $2.646\pm 0.072$ & $2.629\pm 0.068$ & $2.608\pm 0.070$ \\
  $M_B$ & $-19.128\pm 0.058$ & $-19.133\pm 0.066$ & $-19.064\pm 0.057$ & $-19.189\pm 0.055$ \\
  $\Delta_M$ & $-0.052\pm 0.022$ & $-0.058\pm 0.022$ & $-0.056\pm 0.021$ & $-0.059\pm 0.022$ \\
  $\ln \mathcal{L}_{\rm max}$ & 330.591& $331.055$ & $325.141$ & $313.746$\\
  $\Delta {\rm AIC}$ & 0 & 1.072 & 10.900 & 31.690 \\
  $\Delta {\rm BIC}$ & 0 & 5.718 & 10.900 & 27.044 \\
  \hline
\end{tabular}\label{tab:parameter2}
\end{table}
\begin{multicols}{2}

Next, we combine JLA+H(z)+BAO datasets to make an analysis. The results are given in Table \ref{tab:parameter3}. Compared with Table \ref{tab:parameter2}, we can see that adding the BAO data almost does not change the best-fitting parameters of $\Lambda$CDM and $R_h=ct$ models. The most obvious change happens in PL model, in which the Hubble constant is reduced to $h_0=0.645\pm 0.015$ and the power-law index $n$ is reduced to be consistent with 1. The sound horizons in these three models are consistent with each other. Among these models, the PL and $R_h=ct$ models are decisively disfavored compared with $\Lambda$CDM. But there is still no strong evidence to distinguish between flat and non-flat $\Lambda$CDM. The deceleration parameters of flat-$\Lambda$CDM, non-flat $\Lambda$CDM and PL models are $q_0=-0.545\pm0.031$, $-0.424\pm 0.122$ and $-0.080\pm 0.030$, respectively.
\end{multicols}
\begin{table}[htbp]
\centering
\caption{\small{The best-fitting parameters and their $1\sigma$ uncertainties from JLA+H(z)+BAO.}}
\begin{tabular}{crrrr}
  \hline
  & flat $\Lambda$CDM & non-flat $\Lambda$CDM & PL & $R_h=ct$ \\
  \hline
  $h_0$ & $0.688\pm 0.018$ & $0.676\pm 0.019$ & $0.645\pm 0.015$ & $0.625\pm 0.014$ \\
  $\Omega_m$ & $0.303\pm 0.021$ & $0.214\pm 0.059$ & \---- & \---- \\
  $\Omega_\Lambda$ & \---- & $0.531\pm 0.118$ & \---- & \---- \\
  $n$ & \---- & \---- & $1.087\pm 0.035$ & \---- \\
  $\alpha$ & $0.127\pm 0.006$ & $0.127\pm 0.006$ & $0.125\pm 0.006$ & $0.124\pm 0.006$ \\
  $\beta$ & $2.652\pm 0.069$ & $2.646\pm 0.077$ & $2.612\pm 0.062$ & $2.611\pm 0.070$ \\
  $M_B$ & $-19.099\pm 0.055$ & $-19.124\pm 0.055$ & $-19.144\pm 0.053$ & $-19.189\pm 0.054$ \\
  $\Delta_M$ & $-0.054\pm 0.022$ & $-0.052\pm 0.023$ & $-0.060\pm 0.024$ & $-0.062\pm 0.020$ \\
  $r_d$/Mpc & $150.012\pm 3.237$ & $150.028\pm 2.998$ & $150.102\pm 2.570$ & $150.640\pm 2.871$ \\
  $\ln \mathcal{L}_{\rm max}$ & $327.264$ & $328.327$  & $311.807$ & $308.127$\\
  $\Delta {\rm AIC}$ & 0 & -0.126  & 30.914 & 36.274 \\
  $\Delta {\rm BIC}$ & 0 & 4.529  & 30.914 & 31.619 \\
  \hline
\end{tabular}\label{tab:parameter3}
\end{table}
\begin{multicols}{2}

We also check if the H(z) or BAO data alone disfavor any model or not. By using the 30 H(z) data alone, we find that flat and non-flat $\Lambda$CDM models have approximately equal maximum likelihoods, so as PL and $R_h=ct$ models. The best-fitting Hubble constant is $H_0\sim 67~{\rm km~s^{-1}~Mpc^{-1}}$ for flat and non-flat $\Lambda$CDM models, and $H_0\sim 62~{\rm km~s^{-1}~Mpc^{-1}}$ for PL and $R_h=ct$ models. The $\Delta$AIC values are 2.1, 2.1 and 0.1 for non-flat $\Lambda$CDM, PL and $R_h=ct$ models, respectively. The $\Delta$BIC values are 3.5, 2.1 and -1.3 for non-flat $\Lambda$CDM, PL and $R_h=ct$ models, respectively. Therefore, although there is no strong evidence for favoring one model against the others, non-flat $\Lambda$CDM and PL models seem to be marginally disfavored. According to BIC, the H(z) data slightly favor $R_h=ct$ against flat $\Lambda$CDM. Since there are only seven BAO data points, most model parameters couldn't be tightly constrained by BAO data alone, and no model is preferred over the others.

Finally, to avoid model-dependence, we apply the cosmography method, and use SNe Ia data with $z<0.2$ to constrain the deceleration parameter. The best-fitting parameters are $q_0=-0.372\pm 0.181$, $\alpha=0.133\pm 0.008$, $\beta=2.731\pm 0.103$, $M_B=-19.020\pm 0.032$, $\Delta_M=-0.101\pm 0.032$. Since there are only six H(z) data points and one BAO data point at redshift $z<0.2$, adding H(z) and BAO data almost does not change the result. The deceleration parameter is consistent with that of $\Lambda$CDM model within $1\sigma$ uncertainty, and is not zeros at $2\sigma$ confidence level. Since $R_h=ct$ model predicts a null deceleration parameter, it can be ruled out. Note that PL model may have a consistent $q_0$ if the PL index $n\sim 1.6$. However, this PL index is in conflict with that constrained from JLA+H(z) (+BAO). Only $\Lambda$CDM model is consistent with both SNe along and the combined data. Therefore, $\Lambda$CDM is still the best model compared with the rest two models.

\section{Discussions and conclusions}\label{sec:conclusions}

In this paper, we combined the publicly available low-redshift data to constrain $\Lambda$CDM model and its two alternatives, i.e. PL model and $R_h=ct$ model. For $\Lambda$CDM model, we consider flat and non-flat models separately. It is found that, by using the JLA compilation of SNe Ia alone, $R_h=ct$ model is conclusively disfavored against $\Lambda$CDM and PL models. However, it is indistinguishable between $\Lambda$CDM and PL models based on the JLA alone. The power-law index of PL model is about 1.4. This supports that the universe is really accelerating. By using the H(z) or BAO data alone, no model is strongly favored against the others. If we combine JLA and H(z) datasets, there is conclusive evidence disfavor PL and $R_h=ct$ models against $\Lambda$CDM model. Finally, the combined data of JLA+H(z)+BAO also conclusively disfavor PL and $R_h=ct$ models. In addition, the Hubble constant constrained in the $\Lambda$CDM model is consistent with that obtained from the CMB. However, in the PL and $R_h=ct$ models, the Hubble constant is much smaller. Therefore, we conclude that the local probes favor $\Lambda$CDM gainst the rest two models. However, there is no strong evidence to distinguish between flat and non-flat $\Lambda$CDM models.

Shafer \cite{Shafer:2015kda} analyzed two different compilations of SNe Ia and BAO data sets, and found that neither $\Lambda$CDM model nor PL model is strongly preferred over the other if SNe Ia or BAO data are analyzed separately, but the combined analysis of SNe Ia and BAO data strongly prefers $\Lambda$CDM model over PL model. On the other hand, the $R_h=ct$ model is conclusively disfavoured by the SNe alone. Our calculations confirm the results of Ref.\cite{Shafer:2015kda}. By adding the $H(z)$ data to SNe Ia and BAO, we found the significance of disfavoring the PL and $R_h=ct$ models can be highly improved. In addition, we used the cosmography method to constrain the deceleration parameter, and found that only the $\Lambda$CDM model has a consistent deceleration parameters with cosmography.

Tutusaus et al. \cite{Tutusaus:2017ibk} analyzed the similar datasets and found that both $\Lambda$CDM and PL models can fit the local probes equivalently well. The power-law index of PL model they obtained is slightly smaller than 1, so they doubted if the cosmic acceleration is really proven by the local probes. The main difference between Ref.\cite{Tutusaus:2017ibk} and our paper is that, in the former, the authors took into consideration the possible redshift dependence of the absolute luminosity of SNe Ia. They considered four parameterizations of such a dependence, each of which has two parameters, i.e. one amplitude parameter $\epsilon$ and one power-law index $\delta$. To avoid the degeneracy between parameters, they fixed $\delta$ to some arbitrary values. In our paper, we adopted the standard procedure and didn't consider such a dependence. This is because that there is no evidence for such a dependence. Especially, there is no reason why such a dependence, if really exists, can be parameterized in these forms. We tried to constrain $\epsilon$ and $\delta$ with other parameters simultaneously, but found that these two parameters couldn't be tightly constrained. This implies that the parameterizations are not appropriate. It is always possible to eliminate the acceleration if we properly parameterize the evolution term. Riess et al. \cite{Riess:2017lxs} pointed out that the better fit of PL model than $\Lambda$CDM may be due to the small number of SNe at $z>1$ in JLA. With more high-redshift SNe, PL model is no longer as good a fit as $\Lambda$CDM even if the evolution of SNe luminosity is considered.

Recently, the simultaneous detection of gravitational waves (GW) and the electromagnetic counterparts provides another standard siren to test cosmology. The first GW event from binary neutron star merger, GW 170817 \cite{TheLIGOScientific:2017qsa}, was found to be unambiguously associated with a short gamma-ray burst, GRB 170817 \cite{Goldstein:2017mmi,Abbott:2017mdv}. The follow-up observations of this event led to the identification of NGC 4993 as the host galaxy \cite{Coulter:2017}. The advantage of using GW as distance indicator is that it does not relay on other distance ladders and is completely independent of cosmological models. Using the luminosity distance obtained from the GW signals and the redshift of host galaxy, the Hubble constant was tightly constrained to be $70.0^{+12.0}_{-8.0}~{\rm km~s}^{-1}~{\rm Mpc}^{-1}$ \cite{Abbott:2017xzu}. Adding this only one GW data to our JLA+H(z)+BAO sample does not improve the constraint. With the launch of the third-generation GW detectors, such as the Einstein Telescope and the Cosmic Explorer, hundreds to thousands GW events are expected be observed in the next years. We expect that the GW multi-messenger astronomy in the near future will provide deep insight into the universe.

\vspace{2mm}
\acknowledgments{We are grateful to Zhe Chang and Z. C. Zhao for useful discussions.}

\vspace{2mm}
\centerline{\rule{40mm}{0.5pt}}

\end{multicols}

\clearpage
\end{document}